\documentclass[12pt, draftclsnofoot,onecolumn]{IEEEtran}
\usepackage{dsfont}
\usepackage{amssymb}
\usepackage{subfig}
\usepackage{graphicx, amssymb, amsmath}
 \usepackage{epsfig}
\usepackage{extarrows}
\usepackage{color}
\emergencystretch=\maxdimen
\IEEEoverridecommandlockouts
\begin{document}

\title{Energy-Sustainable Traffic Steering for 5G Mobile Networks}

\author{Shan~Zhang,~\IEEEmembership{Member,~IEEE,}
        Ning~Zhang,~\IEEEmembership{Member,~IEEE,}
        Sheng~Zhou,~\IEEEmembership{Member,~IEEE,}
        Jie~Gong,~\IEEEmembership{Member,~IEEE,}
        Zhisheng~Niu,~\IEEEmembership{Fellow,~IEEE,}
        and~Xuemin~(Sherman)~Shen,~\IEEEmembership{Fellow,~IEEE}% <-this % stops a space
\thanks{Shan~Zhang, and Xuemin~(Sherman)~Shen are with the Department of Electrical and Computer Engineering, University of Waterloo, 200 University Avenue West, Waterloo, Ontario, Canada, N2L 3G1 (Email: s327zhan@uwaterloo.ca, xshen@bbcr.uwaterloo.ca).}% <-this % stops a space
\thanks{Ning~Zhang is with the Department of Computing Science, Texas A\&M University-Corpus Christi, 6300 Ocean Dr., Corpus Christi, Texas, USA, 78412 (Email: zhangningbupt@gmail.com).}% <-this % stops a space
\thanks{Sheng~Zhou, and Zhisheng~Niu are with Tsinghua National Laboratory for Information Science and Technology, Tsinghua University, Beijing, 100084, P.R. China (Email: \{sheng.zhou, niuzhs\}@tsinghua.edu.cn).}% <-this % stops a space
\thanks{Jie Gong is with the School of Data and Computer Science, Sun Yat-Sen University, Guangzhou, P.R.China, 510006 (Email: xiaocier04@gmail.com).}% <-this % stops a space
\thanks{This work is sponsored in part by the National Nature Science Foundation of China (No. 91638204, No. 61571265, No. 61461136004), Hitachi R\&D Headquarter, and Natural Sciences and Engineering Research Council of Canada.}% <-this % stops a space 
%\thanks{Part of this work has been presented in Asilomar Conference on Signals, Systems, and Computers~2015 \cite{mine_asilomar}.}
}

\maketitle

\begin{abstract}
Renewable energy harvesting (EH) technology is expected to be pervasively utilized in the next generation (5G) mobile networks to support sustainable network developments and operations.
However, the renewable energy supply is inherently random and intermittent, which could lead to energy outage, energy overflow, quality of service (QoS) degradation, etc.
Accordingly, how to enhance renewable energy sustainability is a critical issue for green networking.
To this end, an energy-sustainable traffic steering framework is proposed in this article, where the traffic load is dynamically adjusted to match with energy distributions in both spatial and temporal domains by means of inter- and intra-tier steering, caching and pushing.
Case studies are carried out, which demonstrate the proposed framework can reduce on-grid energy demand while satisfying QoS requirements.
Research topics and challenges of energy-sustainable traffic steering are also discussed.

\end{abstract}

%\clearpage

\section{Introduction}

%%add more details
The next generation (5G) mobile networks are expected to connect trillions of devices and provide 1000-fold network capacity by 2020 compared with that in 2010.
Network densification (i.e., deploying more small cell base stations (SBSs)) can effectively improve the network capacity through spectrum reuse, and thus is considered as the key cornerstone for 5G.
However, network densification may lead to huge energy consumption, causing heavy burdens to network operators.
To tackle the cumbersome energy consumption, energy harvesting (EH) technology can be leveraged.
Particularly, energy harvesting enabled base stations (EH-BSs) can exploit renewable energy as supplementary or alternative power sources to reduce the operational expenditures (OPEX).
In addition, EH-BSs can be deployed more flexibly without the constraint of power lines.
%Telecommunication equipment manufacturers have designed and built green energy powered off-grid BSs in rural areas.
By 2011, over ten thousands EH-BSs have been deployed globally, and this figure will increase to more than 400,000 by 2020 \cite{EH_num_EHBS_Navigant_report}. % to the Navigant's report

Despite the potential advantages, the inherent randomness of renewable energy poses significant technical challenges to network operations. %\cite{cailing2014sustainability} \cite{EH_BS_sleep_JieGong_TC2014}.
{{Specifically, the mismatch between harvested energy and traffic distributions may result in energy outage and/or energy overflow, degrading the quality of service (QoS) and energy utilization.
Thus, in addition to energy efficiency, a new performance measure ``energy sustainability'' should be introduced to keep the energy outage and the energy overflow probabilities as low as possible \cite{cai11_energy_sustainability_concept}. }}
To this end, we propose a traffic steering framework to enhance energy sustainability in networks with EH-BSs.

Traffic steering goes beyond traffic offloading, which pro-actively adjusts traffic distribution to match with and better utilize network resources, aiming at enhancing network performance or providing better QoS \cite{Nokia_TrafficSteering_WhitePaper}.
The proposed framework encompasses three approaches: inter-tier steering, intra-tier steering, content caching and pushing. 
Specifically, inter-tier steering adjusts the traffic load of each tier according to the variation of renewable energy arrival rate in a large time scale.
In addition, intra-tier steering shifts traffic among neighboring BSs to further grain traffic load in spatial domain, while content caching and pushing reshape temporal traffic load to overcome the small time scale randomness, based on the instant energy status.
Together these three approaches can match traffic demand to renewable energy supply in both spatial and temporal domains, reducing the probabilities of energy outage and overflow.
Therefore, the proposed framework provides two-fold benefits of greenness and QoS provisioning, achieving energy-sustainable networking.

The remainder of this paper is organized as follows. An overview of 5G networks with EH is firstly presented in Section~\ref{sec_HetNet_with EH}. Then, the energy-sustainable traffic steering framework is introduced in Section~\ref{sec_solution}, including detailed methods, research topics and challenges. Case studies are conducted to reveal the effectiveness of energy-sustainable traffic steering in Section~\ref{sec_CaseStudy}, followed by the conclusions. 

%%%%%%%%%%%%%%%%%%%%%%%%%%%%%%%%%%%%%%%%%%%%%%%%%%%%%%%%%%%%%%%%%%%%%%%%%%%%%%%%%%%%%%%%%%%%%%%%%%%
\section{Heterogeneous Networks with Energy Harvesting}
    \label{sec_HetNet_with EH}

    In this section, we introduce energy harvesting enabled networks, including architecture, challenges, existing solutions and limitations.
    
    \subsection{EH-Enabled 5G Network Architecture}
    
\begin{figure}
    	\center
    	% Requires \usepackage{graphicx}
    	\includegraphics[width=3.5in]{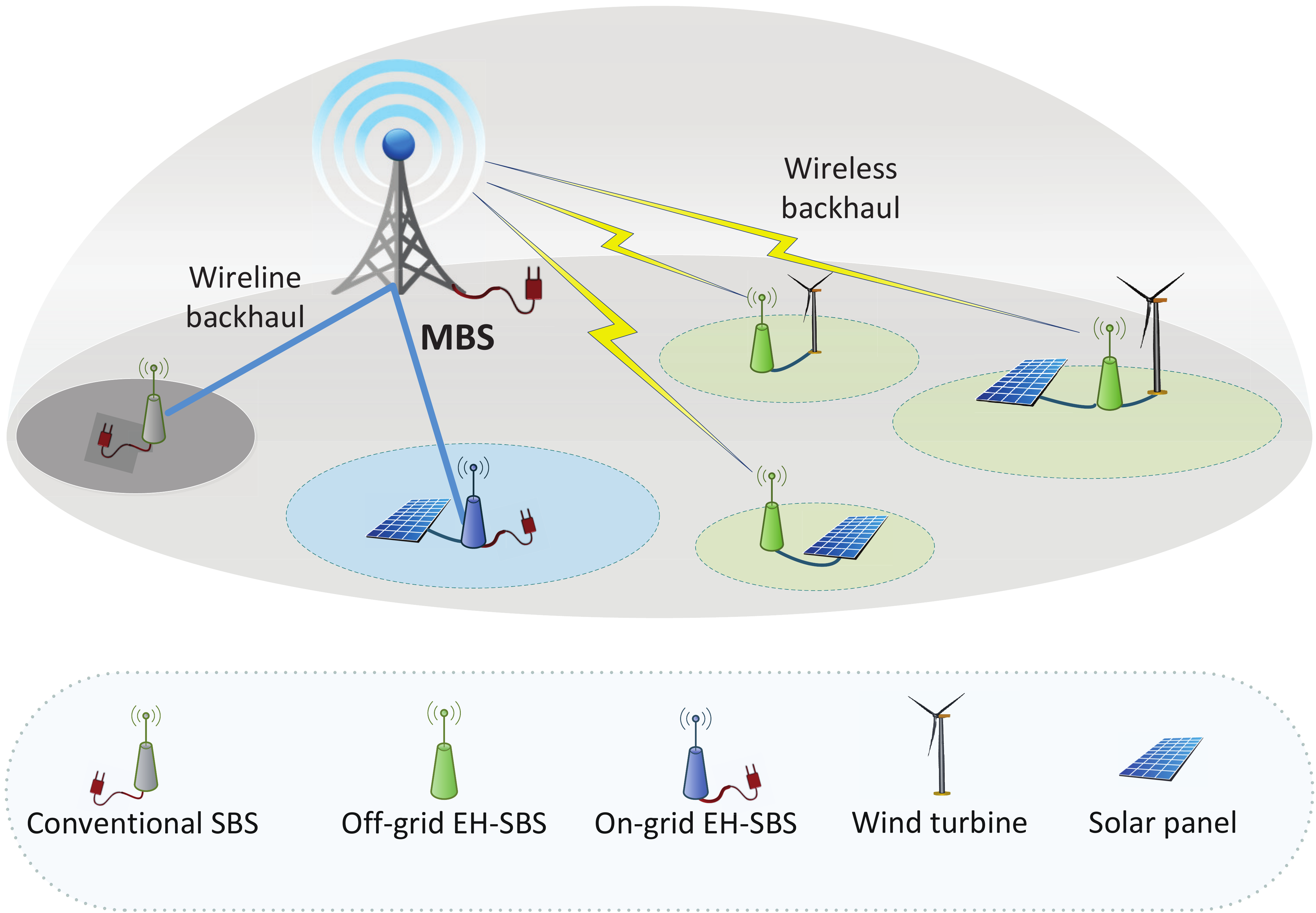}\\
    	\caption{Network architecture with renewable energy harvesting.}\label{fig_HCN_EH}
    \end{figure}

    With the promising EH technologies leveraged, the resulting 5G network architecture is shown as Fig.~\ref{fig_HCN_EH}.
    Particularly, SBSs can be further classified into three types based on the functions and power sources: (1) \textbf{\emph{Off-grid EH-SBSs}}, powered solely by renewable energy without access to power grid; (2) \textbf{\emph{On-grid EH-SBSs}}, powered jointly by power grid and renewable energy; and (3) \textbf{\emph{Conventional SBSs}}, powered only by power grid.
    
    Notice that the three types of SBSs have distinct features.
    Off-grid EH-SBSs enable the most flexible deployment, whereas the QoS cannot be well guaranteed due to the unstable energy supply.
    Thus, off-grid EH-SBSs can be deployed for opportunistic traffic offloading for MBS.
    On-grid EH-SBSs provide reliable services with on-grid power as backup sources, which can also use the renewable energy to reduce the OPEX.
    However, they bring the highest capital expenditure, due to the EH modules and wired connections to power grid.
    Conventional SBSs are the moderate option, which guarantee QoS requirements but with highest on-grid power consumption.
    
    \subsection{Mismatched Traffic and Energy}
    
    As the cellular systems are expected to provide reliable services with guaranteed QoS, the power supply and demand of each BS should be balanced.
    For conventional BSs, the on-grid power supply can be dynamically adjusted based on the traffic variations.
    However, it is much more challenging for EH-BSs, as the renewable energy arrival is usually mismatched with the traffic demand.
    For example, the two-day traffic and renewable energy variations are shown as Fig.~\ref{fig_traffic_energy}, wherein the traffic profile is obtained from real data measurement in EARTH project \cite{EARTH} and the renewable power profile is collected by the Elia group\footnote{{{Elia, Power generation, [Online]. Available: http://www.elia.be/en/grid-data/power-generation, [Accessed: Nov. 2, 2016].}}}. %\cite{Elia_solar_wind_data}
    The mismatch of renewable energy and the traffic load variations may bring following problems:
    %% can be cut here!!
    \subsubsection{{{Renewable Energy Outage}}}
    happens when the energy arrival rate is lower than the BS power demand, which may cause additional on-grid power consumption or degrade the QoS. For example, the off-grid EH-BSs even have to be shut down when the renewable energy is insufficient to support their static power need \cite{EARTH}. Although multiplexing diverse renewable energy sources helps to improve the reliability, the randomness still exists and energy outage can not be avoided.
    \subsubsection{{{Renewable Energy Overflow}}}
    occurs when the renewable energy is oversupplied compared with the traffic demand. To address this problem, batteries can be used to store the redundant energy for future use. {{However, in practical systems, the battery capacity is usually limited due to high cost, and energy overflow is still inevitable.}}
    \subsubsection{{{Spatial Supply-Demand Imbalance}}}
    is due to the diverse energy sources of different BSs. Furthermore, renewable energy is non-uniformly distributed in spatial domain, which may be mismatched with the traffic load in spatial domain. Thus, neighboring BSs can have imbalanced renewable energy supply across the network, leading to inefficient network-wide renewable energy utilization.
    
    \begin{figure}
    	\center
    	\includegraphics[width=3in]{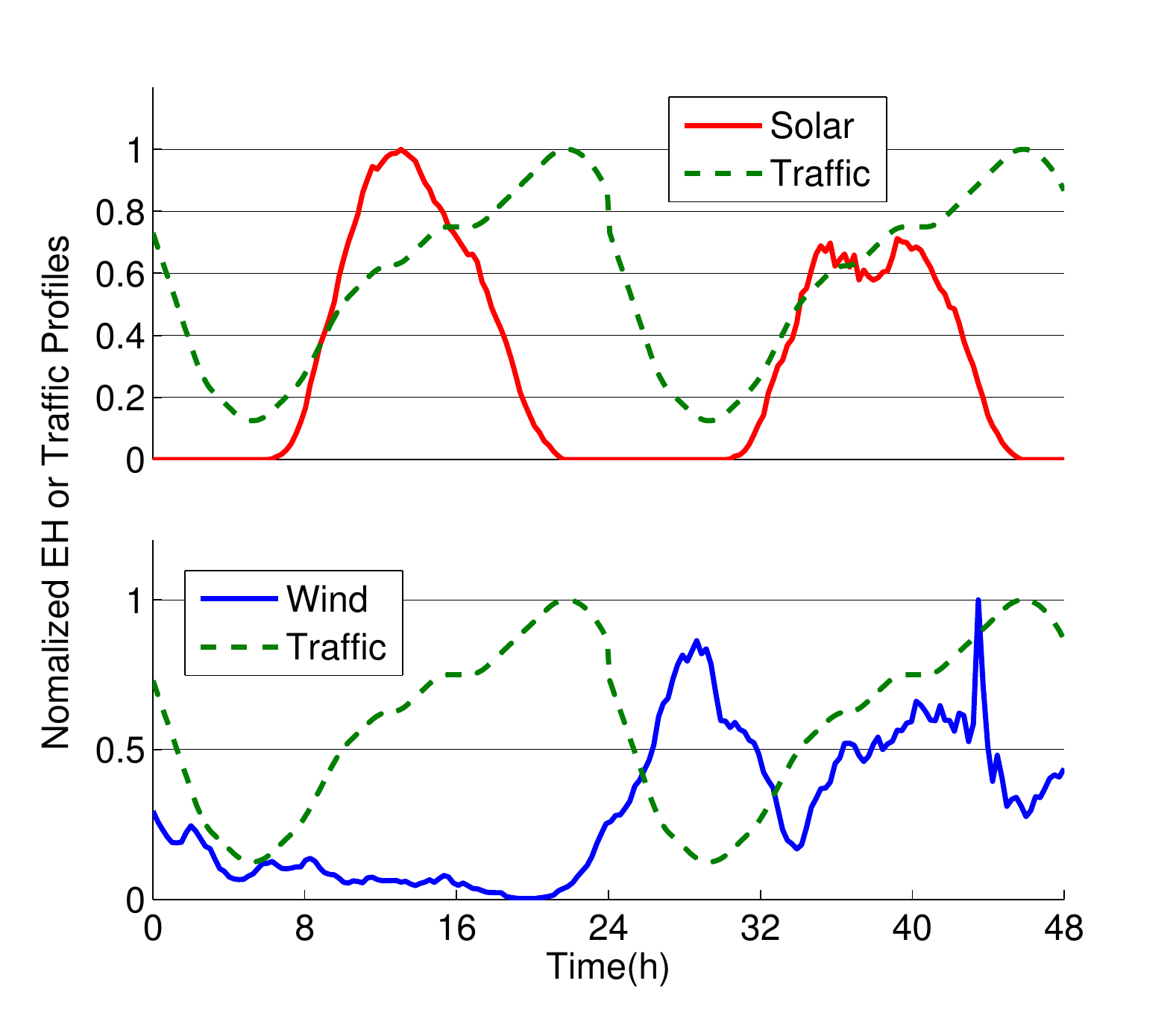}\\
    	\caption{Daily energy harvesting and traffic demand profiles.}\label{fig_traffic_energy}
    \end{figure}
    
    \subsection{Energy-Sustainable Networking}
    
    The challenges of renewable energy dictate that the network design criterion should shift from minimizing the total energy consumption towards energy sustainability, i.e., to sustain traffic while satisfying the QoS requirements with energy dynamics.
    Specifically, energy-sustainable networking improves the utilization of renewable energy to mitigate energy outage and overflow, which enables better QoS provisioning and reduces on-grid power demand.
    To achieve this goal, the energy supply and the traffic demand should be matched with each other in both spatial and temporal domains \cite{zhengzhongming2014sustainable}.
    
    Existing researches mostly focus on the energy management perspective, which reshapes energy supply to match with the given traffic distribution.
    The methods can be mainly classified into two categories:
    \subsubsection{{{BS-Level Energy Allocation}}}
    Through dynamic charging and discharging, the renewable energy can be reallocated in temporal domain to match the time-varying traffic load. For example, effective online and offline energy scheduling schemes have been proposed to minimize the on-grid power consumption of a single EH-BS, assuming infinite battery capacity \cite{EH_single_BS_TWC13}. However, in real systems, the performance of these methods may be degraded due to the battery limitations.  
    \subsubsection{{{Network-Level Energy Cooperation}}}
    Through energy-transfer among EH-BSs, the renewable energy can be redistributed across the network to match with the traffic load, which can further reduce renewable energy waste \cite{EH_energy_coop_2BS_RZhang_TWC2014}. {{However, the EH-BSs need to be connected through either dedicated power lines or smart grids to implement this approach \cite{Bu12_SmartGrid_cellular_TWC}.}} Besides, the two-way power transmission among EH-BSs may cause the loss of renewable energy. 
    
    Notice that the existing methods of energy management manifest their limitations in terms of performance and prerequisite power infrastructure, we seek solutions from the traffic steering perspective, which is more flexible and can be easily realized through control signaling without any additional deployment of power infrastructures. %\cite{mine_GC15_EH}.
    Our previous work was the first to adopt EH-SBS traffic offloading to tackle renewable energy dynamics \cite{mine_JSAC_EH}. 
    In this article, we propose a comprehensive traffic steering framework, where traffic is manipulated dynamically in different spatial- and temporal- scales to match the renewable energy supply.

%%%%%%%%%%%%%%%%%%%%%%%%%%%%%%%%%%%%%%%%%%%%%%%%%%%%%%%%%%%%%%%%%%%%%%%%%%%%%%%%%%%%%%%%%%%%%%%%%%%
\section{Energy-Sustainable Traffic Steering}
    \label{sec_solution}

    In this section, we first introduce the concept and applications of traffic steering, and then propose an energy-sustainable traffic steering framework to realize traffic-energy matching in HetNets with EH.
    
    \subsection{Traffic Steering Concept}
    
    The heterogeneous 5G networks call for a sufficient utilization of available resources to support the dynamic and differentiated traffic demand.
    However, the conventional user association method is mainly based on received signal to interference and noise ratio (SINR), which can fail to meet this requirement.
    To deal with this challenge, traffic steering redistributes traffic load across the network based on radio resources to optimize the performance of networks and end users.
    As user association goes beyond ``SINR-based'' to ``resource-aware'', traffic steering can effectively enhance network utility through appropriate traffic-resource matching.
    
    The objectives and policies of traffic steering can be diverse.
    For example, traffic can be steered from heavily-loaded cells to lightly-loaded ones for load balancing, aiming at maximizing network capacity.
    Besides, in networks with multiple radio access technologies (RATs), users can be steered to different RATs according to their mobility, such that call dropping probability can be reduced. %\cite{TrafficSteering_Munoz2013_CM}
    In addition, traffic steering can be also performed in temporal domain through transmission scheduling and rate control.
    For instance, the transmission of best-effort traffic can be postponed (i.e., steered to future time slots) when the traffic demand exceeds network capacity.
    
    \begin{figure*}
    	\centering
    	\includegraphics[width=4.5in]{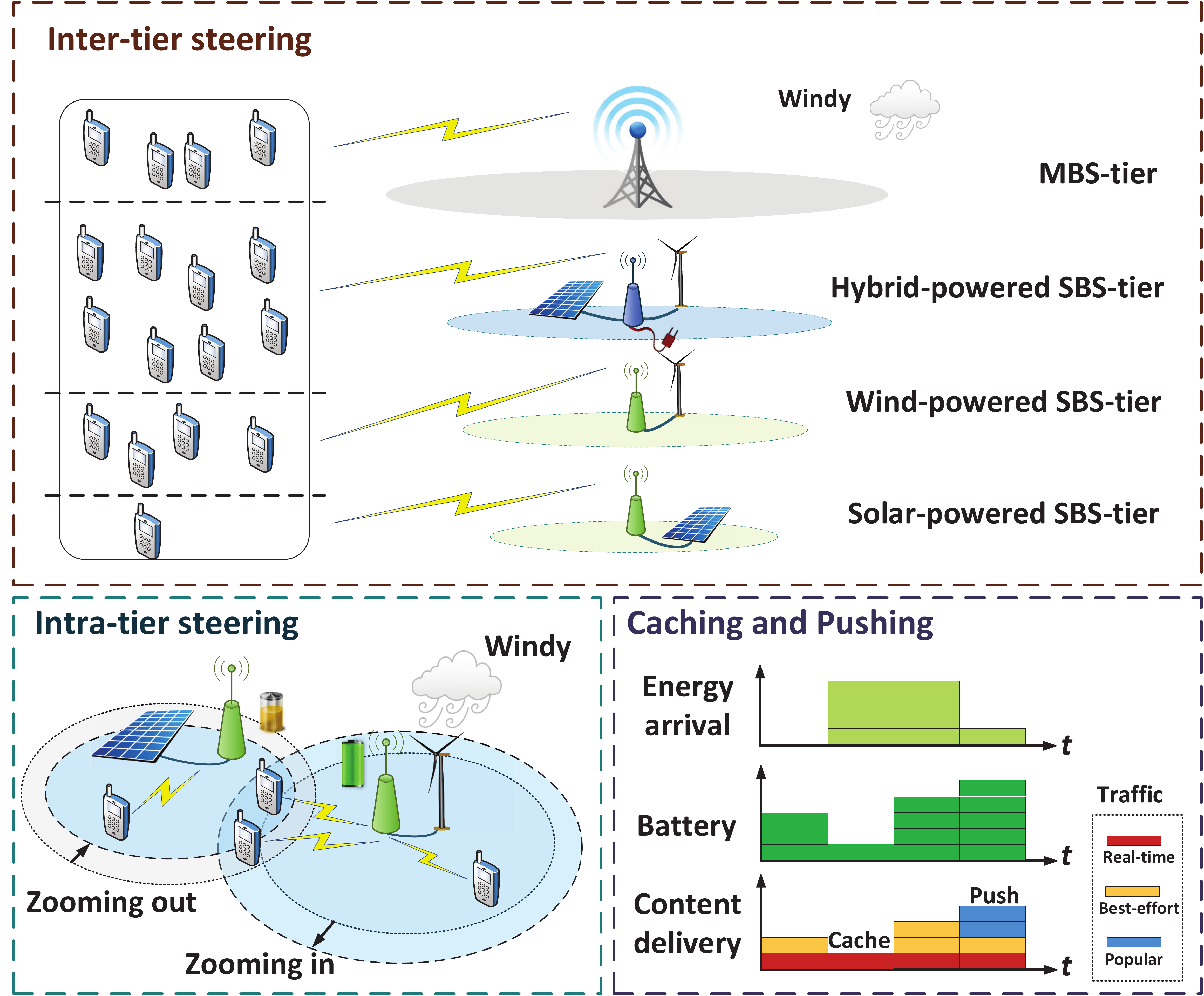}\\
    	\caption{Energy-sustainable traffic steering framework.}\label{fig_matching_example}
    \end{figure*}
    
    \subsection{Traffic Steering with EH}
    
    With EH leveraged, the key challenge is that the service capability of a BS can vary dynamically with renewable energy supply, which requires the traffic load matched with the corresponding service capability to fully utilize renewable energy. 
    To this end, we propose an energy-sustainable traffic steering framework which encompasses three main approaches, as shown in Fig.~\ref{fig_matching_example}.
    Inter-tier steering optimizes the amount of traffic steered to different tiers in time large scale, based on the statistic information of renewable energy supply.
    Then, intra-tier steering, caching and pushing dynamically reshape traffic load in smaller scales, to achieve fine-grained traffic-energy matching.
    Specifically, intra-tier steering adjusts the BS-level traffic load to the corresponding energy supply through the cooperation among multiple neighboring cells.
    Meanwhile, caching and pushing schedule content delivery by exploiting the content information and differentiated QoS requirements, which can be conducted independently at each BS to further deal with small time scale randomness of energy and traffic dynamics.
    
    \subsubsection{\textbf{Inter-Tier Steering}}
    
    As shown in Fig.~\ref{fig_HCN_EH}, BSs can be further divided into multiple tiers based on power sources, cell type and other system parameters.
    Each tier has different service capabilities, which can vary dynamically with renewable energy arrival rate.
    Energy-sustainable inter-tier traffic steering dynamically optimizes the amount of traffic steered to each network tier, based on the information of energy supply and other system parameters.
    Intuitively, more traffic can be steered to the EH-SBS tiers with sufficient energy for service, which reduce the on-grid power consumption of other tiers.
    On the contrary, EH-SBSs with insufficient energy supply can reduce transmit power consumption by serving less users to maintain the power balance (such as the solar-powered SBSs on cloudy day).
    In addition, EH-SBSs can be deactivated when the power supply is even lower (e.g., solar-powered SBSs at midnight), which can further save the static power consumption (such as airconditioner) to enhance renewable energy sustainability. %\cite{mine_TWC_SCoff}.
    
    With intelligent inter-tier traffic steering, the traffic load can be reshaped in both spatial and temporal domains simultaneously.
    In spatial domain, the traffic load of each tier is optimized based on their service capability, as shown in Fig.~\ref{fig_matching_example}.
    From the whole network perspective, the traffic load supported by different energy sources is also dynamically adjusted with on-grid BSs serving as backups, which in fact realize temporal traffic-energy matching.

    \subsubsection{\textbf{Intra-Tier Steering}}
    
    Intra-tier traffic steering further adjusts BS-level traffic load through methods of cell zooming and BS cooperation, which further grain traffic load in spatial domain based on energy status.		
    With cell zooming, the coverage of a BS can be enlarged (i.e., zoom in) or shrink (i.e., zoom out) to adjust the corresponding traffic load, through transmit power control and antenna tilting. %\cite{Cell_zooming_Mag_2010}.
    In conventional on-grid networks, traffic is generally steered from heavily-loaded cells to lightly-loaded ones, leading to load-dependent cell size.				
    With EH technology implemented, EH-BSs should further adjust cell size based on the corresponding energy supply.
    For example, the EH-BSs with oversupplied energy can zoom in to assist the transmission of neighboring BSs by utilizing the redundant harvested energy.
    In reward, some EH-BSs encountering energy shortage can shrink their coverage to reduce their traffic load (i.e., zooming out).
    By doing so, traffic are steered to EH-SBSs with redundant renewable energy, reducing service outage and battery overflow.
    
    In addition to cell zooming, energy-aware cooperative transmission can be also applied.
    As networks becoming ultra dense, a mobile user may be covered and served by multiple BSs simultaneously.
    In this case, the cooperative transmission of these BSs can be optimized based on their energy status in addition to channel condition.
    For instance, the BSs with insufficient renewable energy or large path loss can reduce transmit power or even turn off, while the others enlarge transmit power to guarantee QoS.
    
    \subsubsection{\textbf{Caching and Pushing}}
    
    {{Caching and pushing aim to schedule content delivery in an optimal manner by exploiting the information of contents, user preference, and differentiated QoS requirements, which can reshape traffic load to deal with the small time scale randomness of renewable energy arrival.}}
    
    The transmission of non-realtime traffic can postponed during energy shortage periods, by caching corresponding contents at BS.
    For example, the packet can be delivered with lower transmission rate to reduce energy consumption as long as the given deadline is satisfied, while the transmission of best effort traffic can be delayed until the energy is sufficient.		 	 
    The main idea of proactive pushing is that the EH-BSs can deliver the popular contents (e.g., videos and news) in a multicast manner before user requests, when the energy is sufficient or oversupplied.
    By storing the contents at end user devices, the amount of data transmission can be reduced.
    In fact, the demand of video streaming, currently accounting for over 50\% of wireless traffic and still increasing, is expected to dominant mobile data service. %\cite{cisco2015global}.
    Surprisingly, 10\% of the most popular videos receive nearly 80\% of views. % \cite{video_popularity_2009}
    Therefore, proactive pushing can effectively reduce the future traffic load. % \cite{GreenDelivery_SZhou_2015}.
    
    Through energy-sustainable caching and pushing, the traffic load is equivalently steered from the periods of energy shortage to the periods of energy oversupply.
    This traffic reshaping enables traffic-energy matching in smaller time scales, further enhancing energy utilization.
    
    \subsection{Research Challenges}
    
    Based on the design scales, inter-tier steering, intra-tier steering, caching and pushing can be conducted at hour-, minute-, and second-levels, respectively, providing both coarse- and fine-grained network optimization.
    As such, energy-sustainable traffic steering can match traffic demand to energy supply flexibly in spatial and temporal domains, providing two-fold benefits of greenness and QoS.
    To realize the potential advantages, several research challenges to implement energy-sustainable traffic steering are discussed as follows.
    
    \subsubsection{{Service Capability with EH}}
    
    Sine traffic steering is mainly to match the traffic load to the corresponding service capability from different spatial-temporal scales, a fundamental problem is to analyze the EH-enabled service capability of each BS and network tier.
    In existing literature, the link-level channel capacity has been extensively studied, with an EH-enabled transmitter and a single or multiple receivers \cite{Ozel13_EH_link_capacity_infinite_battery}.
    {{However, network-level analysis can be much more challenging due to factors such as random traffic distribution, user mobility, differentiated QoS requirements, inter-cell interference, and heterogeneous network architecture.}}
    Stochastic geometry can be applied to QoS analysis of large-scale networks, based on the information of traffic distribution and network topologies. 
    For the conventional on-grid HetNets, network capacity has been investigated with respect to different QoS performance metrics, such as coverage probability and user achievable rate.
    Future work should revisit these issues considering the influence of renewable energy supply in both large and small time scales \cite{EH_net_fundamental_TWC2014_dhillon}.
    
    \subsubsection{{Low-Complexity Operation}}
    
    Traffic steering in spatial domain needs the cooperation of multiple cells from the same or different tiers, where the BS activation/deactivation, power control and user association should be jointly optimized based on the instant information of traffic distribution and renewable energy supply.
    The problem is to minimize the long-term on-grid energy consumption subject to QoS requirements and power constraints, which can be formulated as a mix-integer programming problem.
    Existing literature mainly focus on traffic steering optimization for small-scale networks, whereas the operational complexity can increase exponentially with network scale due to the coupled operation of different cells \cite{Li15_RA_JSAC}.
    {{As future mobile networks are expected to be ultra-densely deployed with massive device connections, low-complexity steering schemes is critical for practical implementation.
    		To this end, the BS operations can be decoupled in both temporal and spatial domains.
    		In the temporal domain, the deactivation of BSs can be determined in large time scale based on the statistic information of energy and traffic arrivals, and then each active BS can further schedule the content delivery in small time scale based on the instant states.
    		In the spatial domain, the concept of self-organized network (SON) can be applied to design effective distributed traffic steering schemes, such that BSs can make decisions independently with local information \cite{zhou16_outage_large_scale}.}}
    
    \subsubsection{{Joint Spatial-Temporal Optimization}}
    
    Existing studies have designed spatial traffic steering schemes based on semi-static traffic model \cite{mine_JSAC_EH}, and also proposed dynamic content caching and pushing schemes from a single BS perspective.
    The joint optimization of spatial and temporal traffic steering offers opportunities to further improve network performance.
    In this case, mobile operators have more degree of freedom to reshape traffic load.
    For instance, the oversupplied energy at a BS can be utilized for: 1) zooming in to assist neighboring cells, 2) serve more traffic from upper-tier on-grid MBSs, 3) pushing popular contents to users pro-actively, and 4) storing the redundant energy in battery for future use.
    The Markov Decision Process (MDP) method can be applied to design optimal steering policies through theoretical analysis of small scale networks, which can provide a guideline to practical network operation.
    Furthermore, advanced machine learning technologies can be implemented to devise spatial-temporal traffic steering method for large scale networks, based on the big data analysis in real systems \cite{Wu16_green_big_data} \cite{Chen15_learning_green}.
     
    \subsubsection{{Tradeoff between System and User Performance}}
    
    Although traffic steering helps to improve network performance, the interests of end user devices may be degraded.
    Some end devices may not be connected with the preferred BS after steered in spatial domain, and thereby they have to increase the transmit power to maintain the uplink QoS.
    Accordingly, the battery of end devices may be quickly consumed.
    Moreover, for the temporal traffic steering, proactive pushing and caching also consume the resources of end devices, which may be unacceptable from the perspective of mobile users.
    Existing traffic steering mainly focus on the performance optimization from the network perspective, while the tradeoff between network and user profits has not been well studied.
    For practical implementation, incentive schemes need be designed to compensate the performance degradation of steered users, whereby the agreement can be achieved between network operators and mobile users.
    Such problems can be modeled as two-player or multi-player games, and game theory can be applied to seek equilibriums among players.
    
%%%%%%%%%%%%%%%%%%%%%%%%%%%%%%%%%%%%%%%%%%%%%%%%%%%%%%%%%%%%%%%%%%%%%%%%%%%%%%%%%%%%%%%%%%%%%%%%%%%
\section{Case Study: Inter-Tier Steering}
    \label{sec_CaseStudy}

        \begin{figure*}[!t]
        	\centering
        	\subfloat[On-grid EH-SBS] {\includegraphics[width=2.5in]{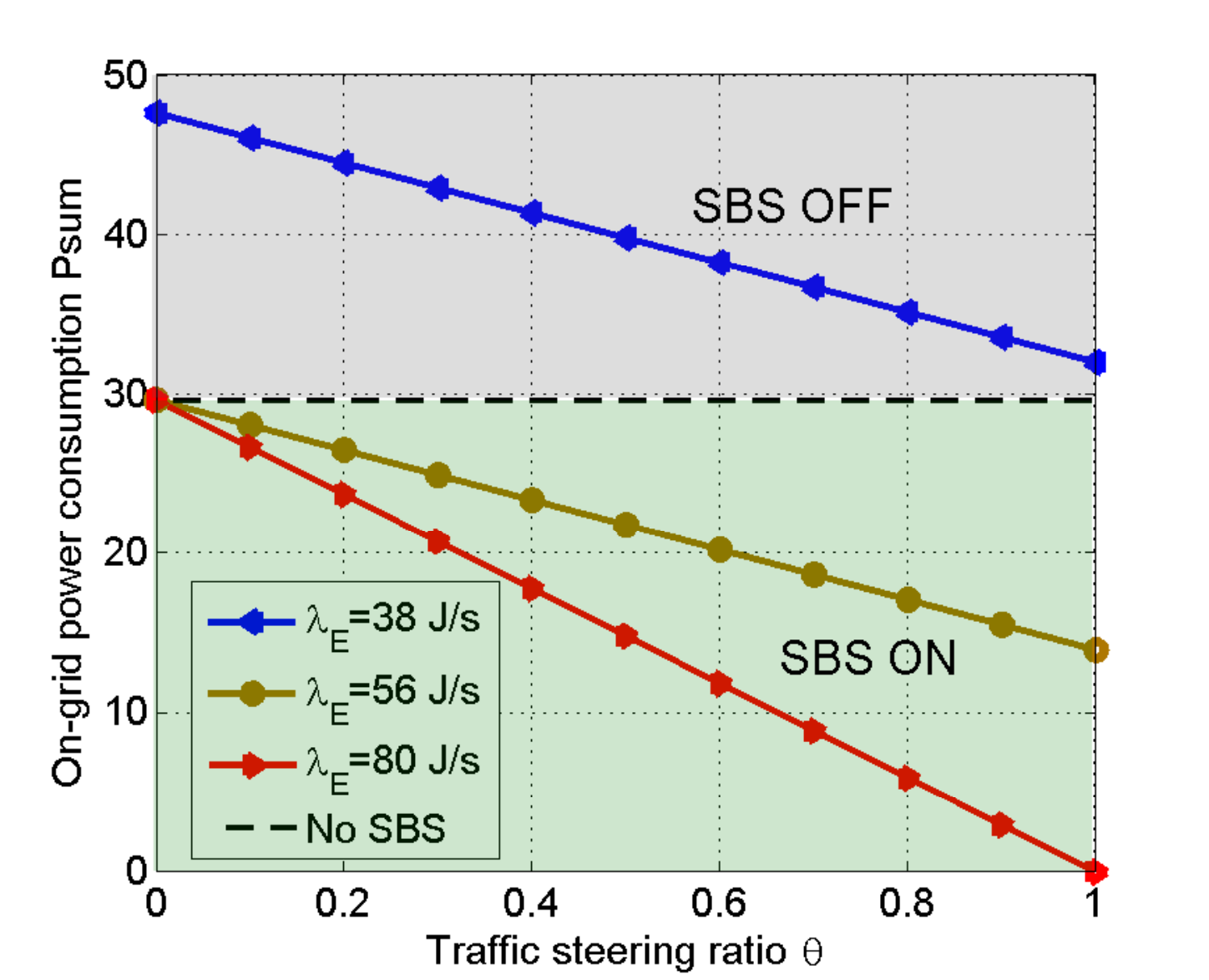}}
        	%\hfil\\
        	\hspace{15mm}
        	\subfloat[Off-grid EH-SBS] {\includegraphics[width=2.5in]{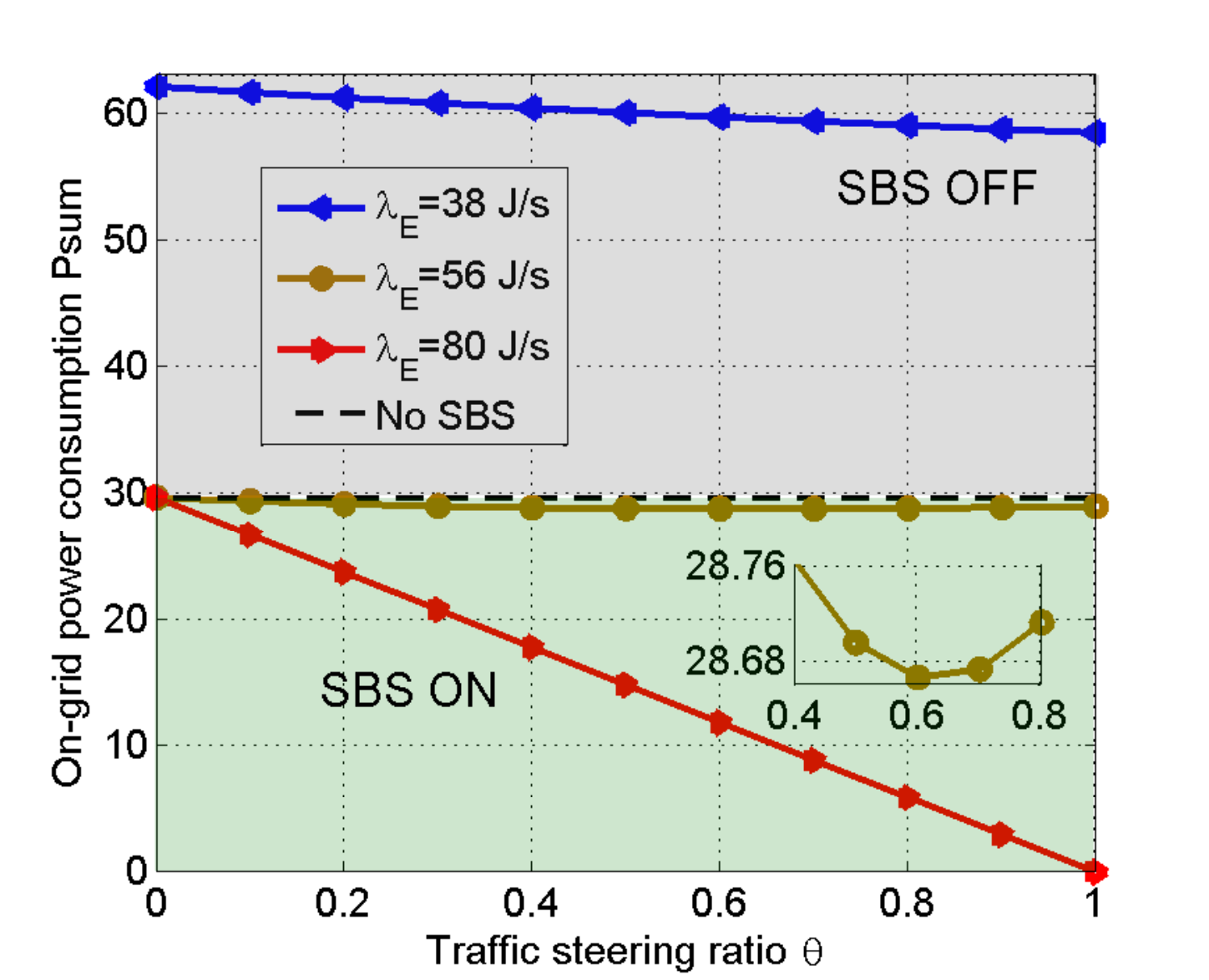}}
        	\caption{Power Consumption with respect to traffic steering ratio.}
        	\label{fig_power}
        \end{figure*}
        
        In this section, a case study is carried out on inter-tier steering to illustrate practical implementation in details.
        We consider a two-tier HetNet consisting of a conventional MBS with radius 1000 m and multiple EH-SBSs with coverage radius 100 m.
        {{The available bandwidths for the MBS and SBS are set as 10 MHz and 2 MHz, respectively.
        		The traffic distribution is modeled as a Poisson Point Process (PPP) with density 1500 /km$^2$, and the average rate per user should be no smaller than 500 kbps.}}
        Traffic within each small cell can be partially or completely steered from the MBS to the EH-SBS depending on the renewable energy arrival rate. 
        Specifically, we analyze how much traffic should be steered to match with the EH-dependent service capability, from the perspective of a typical EH-SBS. 
        Define by $\theta$ the steering ratio, i.e., the percentage of traffic steered to the EH-SBS within the small cell.
        For the given renewable energy arrival rate, we demonstrate the on-grid power consumption variation with respect to the $\theta$, to investigate the influence of energy supply on the BS-level service capability and the optimal traffic steering volume.
        
        Consider the discrete power consumption model, with per unit energy set as 1 Joul.
        The energy arrival at EH-SBS is modeled as Poisson process with rate $\lambda_\mathrm{E}$ for tractable analysis, which is saved at the battery for future use.
        {{Both on-grid and off-grid EH-SBSs are considered, which are assumed be to 500 m away from the MBS.}}
        For the on-grid EH-SBS, it can consume either renewable or on-grid power, but the on-grid power can be only used as backup when the battery is empty.
        For the off-grid EH-SBS, it has to shut down when the battery becomes empty, and meanwhile, all traffic is steered to the MBS through handover procedure, causing additional handover cost.
        {{The BS power consumption model is based on the EARTH project \cite{EARTH}, and power control can be conducted by partially deactivate the subframes.}}
        {{For the wireless channel model, the path loss exponent, noise and interference densities are set as 3.5, -105 dBm/MHz, and -100 dBm/MHz, respectively.}}
        Denote by $P_{\mathrm{sum}}$ the on-grid power required to serve the traffic located within the coverage of small cell.
        Under different energy arrival rate $\lambda_\mathrm{E}$, the relationship between $P_{\mathrm{sum}}$ and the steering ratio $\theta$ is shown in Fig.~\ref{fig_power}, with the handover cost set as 1.5 Joul.
        For comparison, the dash lines shows the power consumption if the EH-SBS is not deployed and all traffic is served by the MBS.
        Therefore, activating EH-SBS for traffic steering even increases power consumption in the regions above the dash lines, and thus the EH-SBSs should be completely deactivated.
        
        %\subsection{On-Grid EH-SBS}
        The on-grid power consumption $P_{\mathrm{sum}}$ is shown to decrease with $\theta$ when the on-grid EH-SBS is active, indicating that the active on-grid EH-SBS should serve traffic as much as possible.
        Furthermore, $P_{\mathrm{sum}}$ decreases with $\lambda_\mathrm{E}$, as more renewable energy can be utilized.
        However, steering traffic to the on-grid EH-SBS does not always save energy, even under the optimal steering ratio.
        For example, the minimal on-grid power consumption is around 32 W when $\lambda_\mathrm{E}$=38 J/s (with $\theta=1$), while only 30 W is needed if the EH-SBS is not activated, shown as the dash lines in Fig.~\ref{fig_power}(a). 
        The reason is due to the tradeoff between transmission and static power consumptions. 
        On the one hand, steering traffic to the SBS helps to reduce transmission power, as the SBS provides higher spectrum efficiency due to short transmission path.
        On the other hand, the active SBS also requires static power independent of data transmission \cite{EARTH}, which may increase the total on-grid power consumption when renewable energy is insufficient.
        This tradeoff is influenced by the renewable energy arrival rate, which directly determines the network power consumption.
        Therefore, the on-grid EH-SBSs should be turned off under low energy arrival rate, whereas they should try to serve as much traffic as possible once activated.
        
        %\subsection{Off-Grid EH-SBS}        
        As for the off-grid EH-SBS, the optimal ON-OFF states also depends on the renewable energy arrival rate.
        However, the optimal traffic steering ratio may not be 1.
        As an example, the energy-optimal steering ratio is 0.6 when the renewable energy arrival rate is 56 W.
        This is due to the tradeoff between the handover cost and the power saved at the MBS.
        With more traffic steered to EH-SBS, the power consumption of the MBS is reduced, whereas the renewable energy of the EH-SBS can be used up more quickly if insufficient, leading to frequent user handover and extra cost.
        This tradeoff determines the optimal steering ratio and ON-OFF status of the off-grid EH-SBS, based on the renewable energy arrival rate.
        %%%%%%%%%%%
        Furthermore, the optimal steering ratio indicates the service capability of EH-SBS, which increases with renewable energy arrival rate.
        When the energy arrival rate is high (e.g., $\lambda_\mathrm{E} = 80$ J/s), the EH-SBS should serve all traffic within coverage (i.e., $\theta=1$), reflecting high service capability.
        On the contrary, EH-SBS should be deactivated and serve no traffic (i.e., $\theta=0$) under low energy arrival rate (e.g., $\lambda_\mathrm{E} = 38$ J/s), reflecting zero service capability.
        
        \begin{figure}[!t]
        	\centering
        	\includegraphics[width=2.5in]{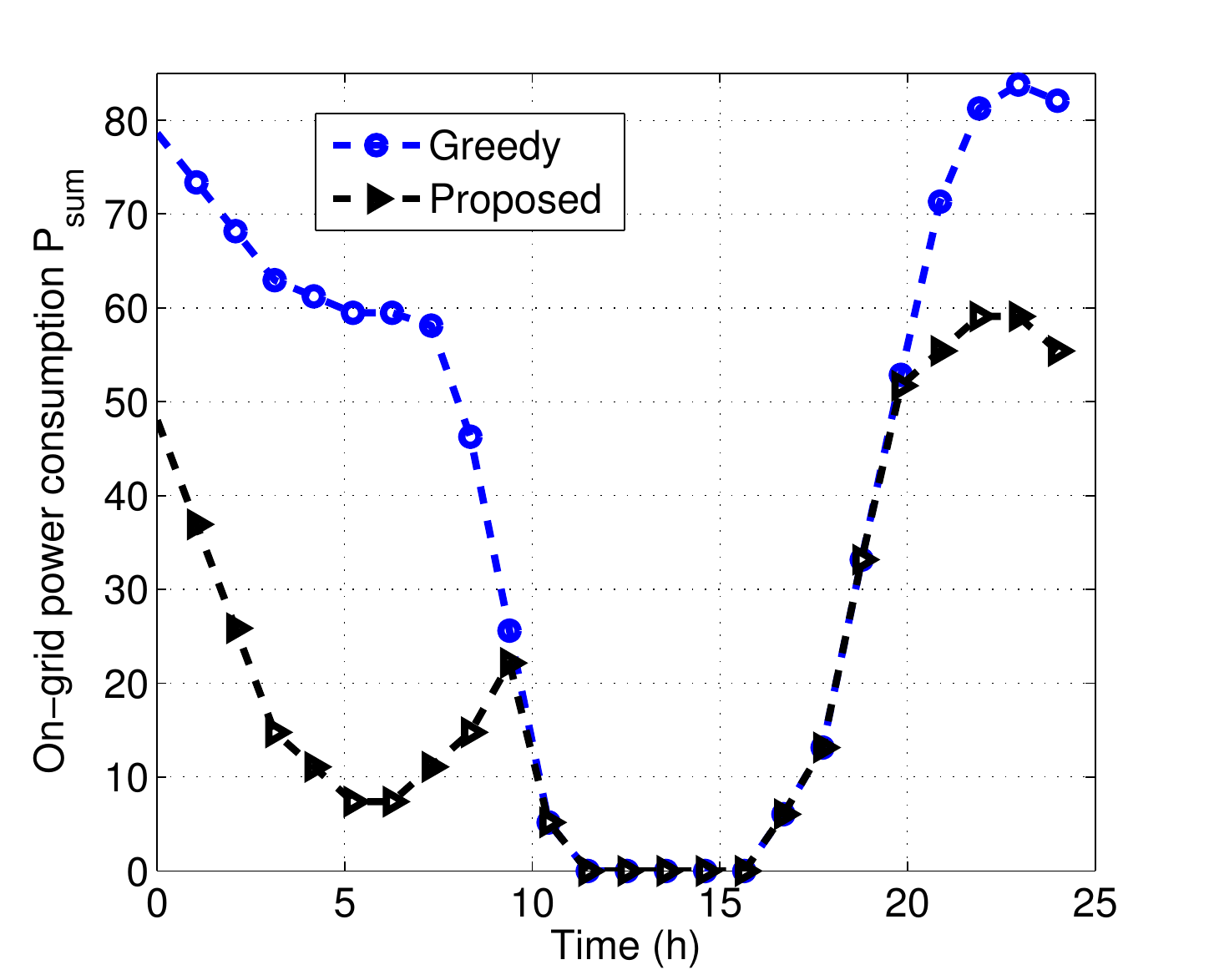}
        	\caption{Daily on-grid power consumptions under different traffic steering schemes.}
        	\label{fig_compare}
        \end{figure}
        
        {{To evaluate the effectiveness of energy-sustainable traffic steering, we illustrate the daily power consumption profiles of an on-grid EH-SBS under different traffic steering schemes. The traffic and solar energy profiles shown in Fig.~\ref{fig_traffic_energy} (the first day) are adopted, with the peak energy arrival rate and traffic load set as 100 /s and 3000 /km$^2$, respectively. The greedy algorithm keeps the EH-SBS on to serve all the traffic within coverage. The proposed energy-sustainable scheme dynamically adjusts the ON/OFF states and the traffic load of the EH-SBSs to minimize the on-grid power consumption, based on the results of Fig.~\ref{fig_power}. The temporal variations of on-grid power consumptions under the two schemes are demonstrated in Fig.~\ref{fig_compare}. Specifically, it is shown that the proposed energy-sustainable traffic steering scheme can reduce the on-grid power consumption by 48\% on average, compared with the greedy scheme. Furthermore, the proposed algorithm is more effective during low energy hours.}}
    
%%%%%%%%%%%%%%%%%%%%%%%%%%%%%%%%%%%%%%%%%%%%%%%%%%%%%%%%%%%%%%%%%%%%%%%%%%%%%%%%%%%%%%%%%%%%%%%%%%%

\section{Conclusions}
    \label{sec_conclusions}
In this article, an energy-sustainable traffic steering framework has been proposed to address the sustainability issue of 5G networks by encompassing three approaches: (1) inter-tier steering, (2) intra-tier steering, and (3) content caching and pushing.
The proposed framework can better balance the power demand and supply of individual EH-BSs, by matching the traffic load to renewable energy distribution in both spatial and temporal domains.
The case study on inter-tier offloading has demonstrated that the ON-OFF status and the steered traffic load of each EH-SBS should be adapted to the corresponding energy-dependent service capability, such that the on-grid power consumption can be effectively reduced.
Future research topics and challenges are also discussed.
%% design & challenges

%%%%%%%%%%%%%%%%%%%%%%%%%%%%%%%%%%%%%%%%%%%%%%%%%%%%%%%%%%%%%%%%%%%%%%%%%%%%%%%%%%%%%%%%%%%%%%%%%%%

%\bibliographystyle{IEEEtran}
%\bibliography{REF}

\begin{thebibliography}{100}
		
	\bibitem{EH_num_EHBS_Navigant_report}
	R.~Martin, ``Nearly 400,000 off-grid mobile telecommunications base stations
	employing renewable or alternative energy sources will be deployed from 2012
	to 2020,'' Navigant Research, Tech. Rep., Feb. 2013, [Online]. Available: {http://www.navigantresearch.com/newsroom/nearly-400000-off-grid-mobile-telecommunications-base-stations-employing-renewable-or-alternative-energy-sources-will-be-deployed-from-2012-to-2020}, [{A}ccessed: on Nov. 4,
	2016].  
	
	\bibitem{cai11_energy_sustainability_concept}
	L.~X. Cai, H.~V. Poor, Y.~Liu, X.~Shen, and J.~W. Mark, ``Dimensioning network
	deployment and resource management in green messh networks,'' \emph{{IEEE}
		Wireless Commun.}, vol.~18, no.~5, pp. 58--65, Oct. 2011.
	
	\bibitem{Nokia_TrafficSteering_WhitePaper}
	Nokia, ``Business aware traffic steering,'' Nokia Corporation, Tech. Rep., Feb.
	2015, [Online]. Available:
	{http://info.networks.nokia.com/Business-aware-traffic-steering-LP.html}, [{A}ccessed: Nov. 4, 2016].
	
	\bibitem{EARTH}
	G.~Auer, O.~Blume, V.~Giannini, I.~Godor, M.~Imran, Y.~Jading, E.~Katranaras,
	M.~Olsson, D.~Sabella, P.~Skillermark \emph{et~al.}, ``D2.3: energy
	efficiency analysis of the reference systems, areas of improvements and
	target breakdown,'' EARTH Project, Tech. Rep., Nov. 2010, [Online]. Available:
	https://www.ict-earth.eu/publications/deliverables/deliverables.html, 
	[{A}ccessed: Nov. 4, 2016].
	
	\bibitem{zhengzhongming2014sustainable}
	Z.~Zheng, X.~Zhang, L.~X. Cai, R.~Zhang, and X.~Shen, ``Sustainable
	communication and networking in two-tier green cellular networks,''
	\emph{{IEEE} Wireless Commun.}, vol.~21, no.~4, pp. 47--53, Aug. 2014.
	
	\bibitem{EH_single_BS_TWC13}
	D.~W.~K. Ng, E.~S. Lo, and R.~Schober, ``Energy-efficient resource allocation
	in {OFDM} systems with hybrid energy harvesting base station,'' \emph{{IEEE}
		Trans. Wireless Commun.}, vol.~12, no.~7, pp. 3412--3427, Jul. 2013.
	
	\bibitem{EH_energy_coop_2BS_RZhang_TWC2014}
	Y.-K. Chia, S.~Sun, and R.~Zhang, ``Energy cooperation in cellular networks
	with renewable powered base stations,'' \emph{{IEEE} Trans. Wireless
		Commun.}, vol.~13, no.~12, pp. 6996--7010, Dec. 2014.
	
	\bibitem{Bu12_SmartGrid_cellular_TWC}
	S.~Bu, F.~R. Yu, Y.~Cai, and X.~P. Liu, ``When the smart grid meets
	energy-efficient communications: Green wireless cellular networks powered by
	the smart grid,'' \emph{{IEEE} Trans. Wireless Commun.}, vol.~11, no.~8, pp.
	3014--3024, Aug. 2012.
	
	\bibitem{mine_JSAC_EH}
	S.~Zhang, N.~Zhang, S.~Zhou, J.~Gong, Z.~Niu, and X.~Shen, ``Energy-aware
	traffic offloading for green heterogeneous networks,'' \emph{{IEEE} J. Sel.
		Areas Commun.}, vol.~34, no.~5, pp. 1116--1129, May 2016.
		
	\bibitem{Ozel13_EH_link_capacity_infinite_battery}
	O.~Ozel and S.~Ulukus, ``Achieving {AWGN} capacity under stochastic energy
	harvesting,'' \emph{{IEEE} Trans. Inf. Theory.}, vol.~58, no.~10, pp.
	6471--6483, Oct. 2012.
		
	\bibitem{EH_net_fundamental_TWC2014_dhillon}
		H.~S. Dhillon, Y.~Li, P.~Nuggehalli, Z.~Pi, and J.~G. Andrews, ``Fundamentals
		of heterogeneous cellular networks with energy harvesting,'' \emph{{IEEE}
			Trans. Wireless Commun.}, vol.~13, no.~5, pp. 2782--2797, May 2014.
	
	\bibitem{Li15_RA_JSAC} P.~Li, S.~Guo, W.~Zhuang, and B.~Ye, ``On efficient resource allocation for cognitive and cooperative communications,'' \emph{{IEEE} J. Sel. Areas Commun.}, vol.~32, issue.~2, pp. 264--273, Feb. 2014.
	
	\bibitem{zhou16_outage_large_scale} Y.~Zhou, and W.~Zhuang, ``Performance analysis of cooperative communication in decentralized wireless networks with unsaturated traffic,'' \emph{{IEEE}
		Trans. Wireless Commun.}, vol.~15, no.~5, pp. 3518--3530, May 2016.
	
	\bibitem{Wu16_green_big_data} J.~Wu, S.~Guo, J.~Li, and D.~Zeng, ``Big data meet green challenges: Big data toward green applications,'' \emph{{IEEE}
		Syst. J.}, vol.~10, no.~3, pp. 888--900, Sept. 2016.
	
	\bibitem{Chen15_learning_green} X.~Chen, J.~Wu, Y.~Cai, H.~Zhang, and T.~Chen, ``Energy-efficiency oriented traffic offloading in wireless networks: A brief survey and a learning approach for heterogeneous cellular networks,'' \emph{{IEEE} J. Sel. Areas Commun.}, vol.~33, issue.~4, pp. 627--640, April 2015.

\end{thebibliography}

\begin{IEEEbiographynophoto}{\bf Shan Zhang} [M] (s372zhan@uwaterloo.ca) received her Ph.D. degree in Department of Electronic Engineering from Tsinghua University, Beijing, China, in 2016. She is currently a postdoctoral fellow in Department of Electronical and Computer Engineering, University of Waterloo, Ontario, Canada. Her research interests include resource and traffic management for green communication, intelligent vehicular networking, and software defined networking. Dr. Zhang received the Best Paper Award at the Asia-Pacific Conference on Communication in 2013.
\end{IEEEbiographynophoto}
%\vspace{-20mm}

\begin{IEEEbiographynophoto}{\bf Ning Zhang} [M] (zhangningbupt@gmail.com) received the Ph.D degree from University of Waterloo in 2015. He is now an assistant professor in the Department of Computing Science at Texas A\&M University-Corpus Christi. Before that, he was a postdoctoral research fellow at BBCR lab in University of Waterloo. He was the co-recipient of the Best Paper Award at IEEE GLOBECOM 2014 and IEEE WCSP 2015. His current research interests include next generation wireless networks, software defined networking, vehicular networks, and physical layer security.​
\end{IEEEbiographynophoto}
%\vspace{-20mm}

\begin{IEEEbiographynophoto}{\bf Sheng Zhou} [M] (sheng.zhou@tsinghua.edu.cn) received his B.S. and Ph.D. degrees in Electronic Engineering from Tsinghua University, China, in 2005 and 2011, respectively. He is currently an associate professor of Electronic Engineering Department, Tsinghua University. From January to June 2010, he was a visiting student at Wireless System Lab, Electrical Engineering Department, Stanford University, CA, USA. His research interests include cross-layer design for multiple antenna systems, cooperative transmission in cellular systems, and green wireless communications.
\end{IEEEbiographynophoto}

%\vspace{-20mm}

\begin{IEEEbiographynophoto}{\bf Jie Gong} [M] (xiaocier@gmail.com) received his B.S. and Ph.D. degrees in Department of Electronic Engineering in Tsinghua University, Beijing, China, in 2008 and 2013, respectively. He is currently an associate research fellow in School of Data and Computer Science, Sun Yat-sen University, Guangzhou, Guangdong Province, China. He was a co-recipient of the Best Paper Award from IEEE Communications Society Asia-Pacific Board in 2013. His research interests include Cloud RAN, energy harvesting and green wireless communications.
\end{IEEEbiographynophoto}
%\vspace{-20mm}

\begin{IEEEbiographynophoto}{\bf Zhisheng Niu} [F] (niuzhs@tsinghua.edu.cn) graduated from Beijing Jiaotong University, China, in 1985, and got his M.E. and D.E. degrees from Toyohashi University of Technology, Japan, in 1989 and 1992, respectively.  During 1992-94, he worked for Fujitsu Laboratories Ltd., Japan, and in 1994 joined with Tsinghua University, Beijing, China, where he is now a professor at the Department of Electronic Engineering. He is also a guest chair professor of Shandong University, China. His major research interests include queueing theory, traffic engineering, mobile Internet, radio resource management of wireless networks, and green communication and networks.
\end{IEEEbiographynophoto}
%\vspace{-20mm}

\begin{IEEEbiographynophoto}{\bf Xuemin (Sherman) Shen} [F] (xshen@bbcr.uwaterloo.ca) is a university professor, Department of Electrical and Computer Engineering, University of Waterloo, Canada. He is also the associate chair for graduate studies. His research focuses on resource management, wireless network security, social networks, smart grid, and vehicular Ad Hoc and sensor networks. He was an elected member of the IEEE ComSoc Board of Governor, and the chair of the Distinguished Lecturers Selection Committee. He has served as the Technical Program committee chair/co-chair for IEEE Globecom’16, Infocom’14, IEEE VTC’10 Fall, and Globecom’07. He received the Excellent Graduate Supervision Award in 2006, and the Outstanding Performance Award in 2004, 2007, 2010, and 2014 from the University of Waterloo. Professor Shen is a registered professional engineer of Ontario, Canada, an IEEE Fellow, an Engineering Institute of Canada Fellow, a Canadian Academy of Engineering Fellow, a Royal Society of Canada Fellow, and was a Distinguished Lecturer of the IEEE Vehicular Technology Society and the IEEE Communications Society.
\end{IEEEbiographynophoto}

\end{document}